\documentclass[a4paper]{jacow}

\ifboolexpr{bool{xetex} or bool{luatex}}
 {}
 {\usepackage[utf8]{inputenc}}

\usepackage{subfig}
\usepackage{amsmath}
\usepackage{graphics}
\usepackage{amssymb}
 \usepackage{graphicx}
\usepackage{marginnote}
\usepackage{float}
\usepackage{lipsum}

\ifboolexpr{bool{jacowbiblatex}}
 {%
  \addbibresource{jacow-test.bib}
  \addbibresource{biblatex-examples.bib}
 }{}

\copyrightspace[2cm] 

\begin{document}

\title{Emittance Growth in a Plasma Wakefield Accelerator}

\author{O. Mete \thanks{oznur.mete@manchester.ac.uk }, K. Hanahoe, G. Xia, The University of Manchester, Manchester, UK\\
The Cockcroft Institute, Sci-Tech Daresbury, Warrington, UK\\
M. Labiche, Nuclear Physics Group, STFC Daresbury Laboratory, Sci-Tech Daresbury, Warrington, UK}

\maketitle

\begin{abstract}
The interaction of the witness beam with the surrounding plasma particles and wakefields was studied. The implications of the elastic scattering process on beam emittance and, emittance evolution under the focusing and acceleration provided by plasma wakefields were discussed. Simulations results from GEANT4 are presented in this paper.
\end{abstract}

\section{Introduction} 
The next generation of particle physics colliders will need to supplement pp collisions with $e^+e^-$ and ep collisions to deliver precision and to address QCD research needs. Generally each successor collider should push the limits of the energy frontier further.

Plasma accelerators have made tremendous progress in the last few decades since the inception of the idea from Tajima and Dawson \cite{Tajima}. Nowadays, laser wakefield accelerators (LWFA) can achieve MeV-GeV level, electrons through millimetre to centimetre plasma cells \cite{Mangles, Geddes, Leemans1, Leemans2, Wang}. Electron beam driven plasma wakefield acceleration (PWFA) has demonstrated energy doubling for an ultra relativistic $42\,$GeV electron beam in a metre long plasma structure \cite{Blumenfeld}. The accelerating gradients measured in these experiments can be in the range of $10$-$100\,$GeV/m, which are $3$-$4$ orders of magnitude larger than that in today's conventional RF-based particle accelerators. 

Towards the realisation of a collider scheme based on plasma wakefield acceleration, challenges and issues must be explored.
\section{Emittance growth due to Coulomb scattering}
Under the conditions where a beam travels in the vacuum with a constant acceleration, emittance decreases with increasing energy according to the conservation of the area in the phase space given by Louville's theorem. This phenomenon is known as adiabatic damping. However, if the particles in the beam encounter a medium of gas or plasma, emittance diffusion occurs through scattering and competes against the adiabatic damping as suggested in Eq.\ref{eqn:diffusion_eq} \cite{base_theory}; 
\begin{equation}
\Delta \epsilon = \frac{F}{2\gamma\prime}\bigg[\sqrt{\gamma_f}-\sqrt{\gamma_i}\bigg],
\label{eqn:diffusion_eq}
\end{equation}
where $\gamma\prime$ is the rate of change of the acceleration, $\gamma_f$ and $\gamma_i$ are the final and initial beam energies, respectively. F is written as,
\begin{equation}
F = 2 \pi r_e^2 n \bigg[ \frac{-\pi\sigma_0^2mc^2}{\lambda_p e E_{z0}cos(\phi)} \bigg]^{1/2}ln\bigg( \frac{\lambda_D}{R} \bigg),
\label{eqn:F}
\end{equation}
where $r_e$ is the classical electron radius, $m$ is the mass of the electron; $n$ is the number of scattering centres, $\sigma_0$ is the initial beam size interacting with the scattering medium. The constant accelerating field is given as $E_{z0}sin(\phi)$. 

Minimum and maximum scattering angles are determined through uncertainty in the momentum of the incident particle, $p$, (Eq.\ref{eqn:uncertainty}) and the impact parameter, $b$. 
\begin{equation}
\theta = \frac{\Delta p}{p},
\label{eqn:uncertainty}
\end{equation} 
the quantum mechanical limit $\Delta p\,b \geq \hbar$ applies resulting in Eq.\ref{eqn:angles}, where $\hbar$ is the reduced Planck constant,
\begin{equation}
\theta_{min,\,max} = \frac{\hbar}{p\,b_{max,min}}.
\label{eqn:angles}
\end{equation} 
The maximum impact parameter, $b_{max}$, comes from the shielding effect of the atomic electrons for a linear plasma wakefield. In a fully ionised plasma this will correspond to the Debye length, shown in Eq.\ref{eqn:lambdaD}, where $\epsilon_0$ is the electric permitivity of vacuum, $k_B$ is the Boltzmann constant, $T_e$ is the temperature of the plasma electrons, $n_e$ is the number of electrons in the plasma and $e$ is the charge of an electron: 
\begin{equation}
\lambda_D = \sqrt{\frac{\epsilon_0 k_B T_e}{n_e e^2}}.
\label{eqn:lambdaD}
\end{equation}
The Debye length is the distance over which the potential of the nucleus is reduced to $1/e$ of its maximum value within a plasma due to the screening effect of the surrounding plasma electrons. For the nonlinear (bubble or blow-out) regime, where number density of the drive bunch is larger than the plasma, the maximum impact factor corresponds to the radius of the ion cavity \cite{Kirby}.

The minimum impact parameter can be related to the effective Coulomb radius of the nucleus, $R$.  The extrema of the scattering angle can be rewritten as in Eq.\ref{eqn:angles_summary}.
\begin{equation}
\theta_{min} = \frac{\hbar}{p \lambda_D},\,\theta_{max} = \frac{\hbar}{p R}
\label{eqn:angles_summary}
\end{equation}
 \section{Monte Carlo simulations}
\begin{figure}[htpb!]  
\centering
\includegraphics[width=0.45\textwidth] {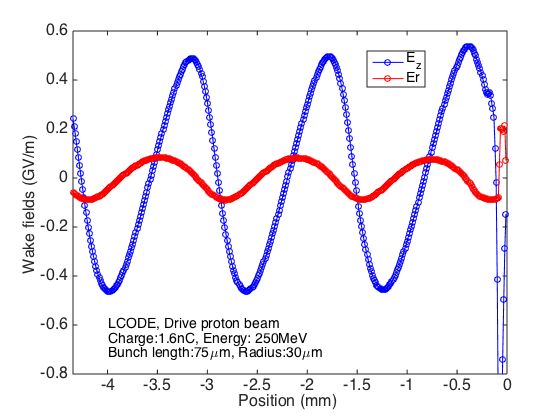}  \\[-0.5ex] 
\caption{Longitudinal and radial plasma wake fields due to a proton bunch that is located at zero.}
\label{fig:wakefields}
\end{figure}

The above theory can be examined by comparing it with the results from a Monte Carlo code which can simulate the particle-matter interactions such as GEANT4 \cite{geant4_1,geant4_2}. A particular scenario was simulated where an electron beam with given parameters, under constant acceleration and focusing, travels through a defined gas column undergoing only elastic Coulomb scattering. An example wake field of a 250 MeV proton drive beam is shown in Fig.\ref{fig:wakefields} within the simulation window of a few plasma periods. This result from LCODE \cite{LCODE} simulations suggests a $0.5\,GV/m$ longitudinal field accompanied by a $0.1\,GV/m$ radial field both of which were implemented in the GEANT4 model. 

\begin{figure}[htpb!]  
\centering
\includegraphics[width=0.4\textwidth] {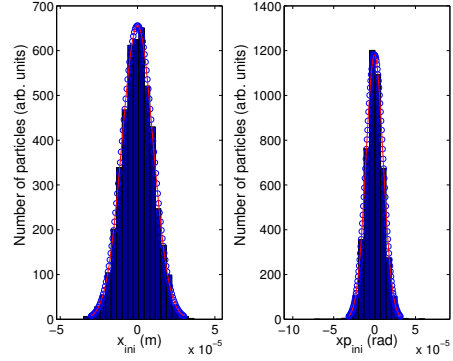}  \\[-0.5ex] 
\caption{Initial particle position (left) and angle (right) distributions provided for the simulations.}
\label{fig:histos1}
\end{figure}

 The initial beam consisting of $10k$ electrons at $10\,$GeV was generated with randomly assigned positions and angles within Gaussian distributions with one standard deviation of $10\,\mu$m, $10\,\mu$rad, respectively, as shown in Fig. \ref{fig:histos1}. These initial values were chosen considering a realistic emittance of an electron beam of $10\,$GeV. 
 
Particles were tracked $500\,$m through neutral Lithium$\,$(Li) gas  ($Z=3$, $a=6.941\,$g/mole) with a density of  $6\times10^{14}$cm$^{-3}$. Li gas was chosen due to its orders of magnitude low scattering cross section compared to the other candidate media such as Rb ($Z=37$). In reality, plasma is produced by ionisation of a channel through a chamber filled with a given gas with a radius given by the ionisation laser specifications \cite{Oz}. Therefore particles travelling through the centre of the chamber may interact with the plasma ions and electrons as well as the surrounding neutral gas when they are scattered out of the plasma channel. With the current technology a plasma column on the order of a mm is possible to produce. Nevertheless, in these simulation studies an arbitrary plasma radius of $100\,$mm was chosen to provide enough clearance for the particles. 

This $500\,$m plasma column with $100\,$mm radius was split into logical sections of $20\,$m, and the transverse phase space of the beam was reconstructed at the end of each section. For the analysis, primary particle tracks were isolated from possible secondary particles.
\subsection{Reconstruction of particle angles and phase space}
\begin{figure}[htpb!]
\centering
\includegraphics[width=0.4\textwidth] {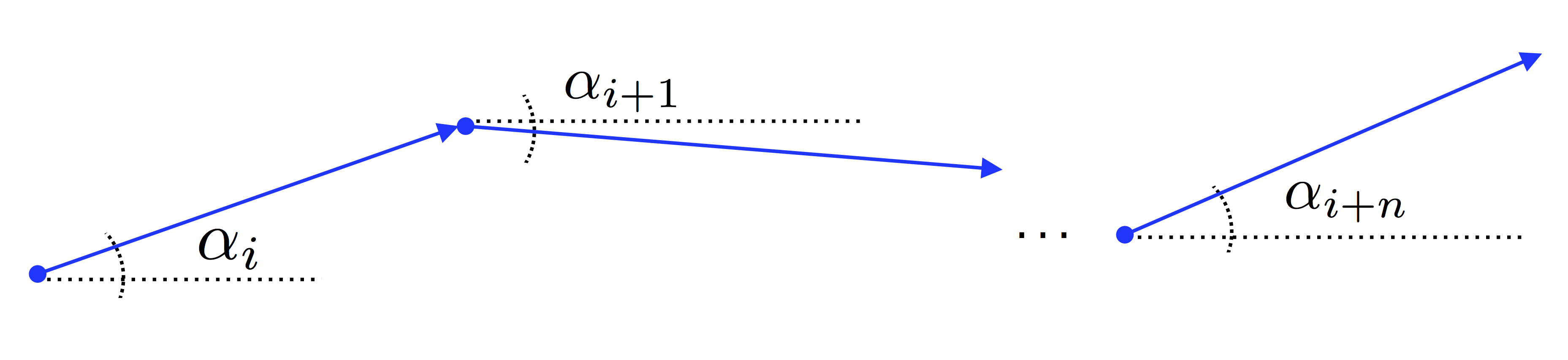} \\ [-0.5ex]
\caption{The mean angle of each particle can be determined for the last $n$ scattering events before the end of each section.}
\label{fig:sc}
\end{figure}

\begin{figure}[htpb!]  
\centering
\includegraphics[width=0.4\textwidth] {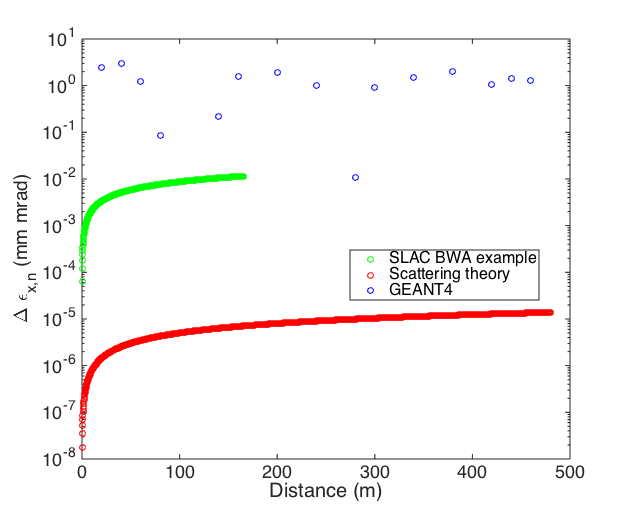}  \\[-0.5ex] 
\caption{The evolution of emittance growth throughout a gas column of $500\,$m according to the theoretical predictions and GEANT4 results.}
\label{fig:emitt_sim}
\end{figure}

In order to produce a transverse phase space distribution after each section, angles of each primary particle should be determined at the end of the section under study. These can be calculated as mean angles corresponding to the last $n$ scattering events as depicted in Fig. \ref{fig:sc}. 
\begin{figure}[htpb!]  
\centering
\subfloat[]{\includegraphics[width=0.4\textwidth] {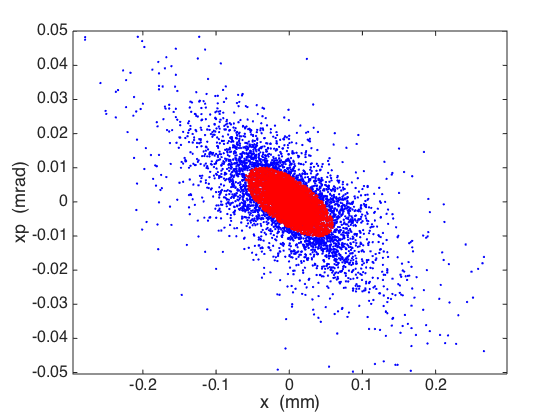}}  \\[-0.5ex] 
\subfloat[]{\includegraphics[width=0.4\textwidth] {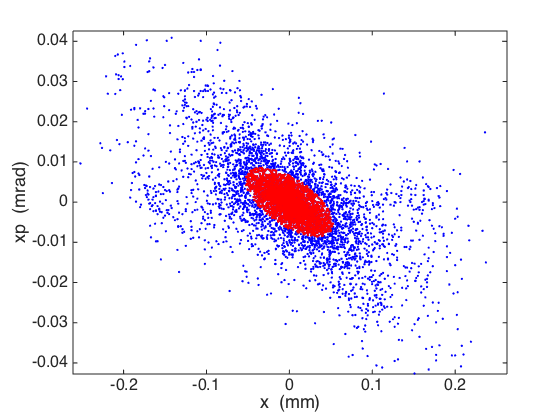}}\\[-0.5ex] 
\caption{Reconstructed phase spaces at (a) $100\,$m and (b) $500\,$m during the beam passage through a Li gas column. Particles within the $1\sigma$ of  the Courant-Snyder ellipses are shown in red.}
\label{fig:phase_spaces}
\end{figure}
Two example transverse phase space distributions, that are constructed from these tracking data, are presented in Figure \ref{fig:phase_spaces}. A geometric cut was implemented to reject the particle tracks falling outside of the plasma column with a radius of $100\,mm$. The rms emittance values were calculated (Eq.\ref{eqn:emitt}) and Courant-Snyder parameters were extracted. The subsequent terms of Eq.\ref{eqn:emitt} are implemented as in Eq.\ref{eqn:term1}, \ref{eqn:term2}, \ref{eqn:term3}, respectively,
\begin{equation}
\epsilon = \sqrt{\langle x^2 \rangle \langle x'^2 \rangle - \langle xx' \rangle^2 }
\label{eqn:emitt}
\end{equation} 
\begin{equation}
\langle x^2 \rangle=\sqrt{\frac{1}{N} \sum_{i=1}^N (x_i - \langle x \rangle)^2}
\label{eqn:term1}
\end{equation} 
\begin{equation}
\langle x'^2 \rangle=\sqrt{\frac{1}{N} \sum_{i=1}^N (x_i' - \langle x' \rangle)^2}
\label{eqn:term2}
\end{equation} 
\begin{equation}
\langle xx' \rangle = \sum_{i=1}^N (x_i - \langle x \rangle )(x_i' - \langle x' \rangle )
\label{eqn:term3}
\end{equation} 
where $N$ is the number of particles considered in the calculation, $x_i$ and $x_i'$ are the coordinates in the phase space. The evolution of the normalised emittance growth in the gaseous media and theoretical prediction are presented in Fig. \ref{fig:emitt_sim}. Green curve is the growth predicted in \cite{base_theory} whereas red curve represents the same model with the beam parameters used in this study. Theoretical emittance growth values were calculated for the a Debye length of $0.3\,\mu$m  assuming a $1\,eV$ of excess energy for the plasma electrons at the given particle density. The GEANT4 results are given in blue dots. Accordingly, at 480m the emittance growth due to collisions is $1.1\times10^{-5}$ of the total emittance of $1.3\,$mm$\,$mrad. This is only a fraction of the total beam emittance. Another cause of emittance growth in a plasma accelerator can be mismatching between radial plasma focusing and the beam beta function. 
\section{Matching beam and plasma}
Beam envelope can undergo betatron oscillations due to the focusing component of the wakefields, $E_r$. Beam emittance can be minimised when plasma focusing term, $K=eE_r/rm_e\gamma c^2$, compensates against beam divergence by satisfying $K\approx\epsilon^2/\sigma_{0}^4$ \cite{unmatched, matched}. This oscillations are currently under study in the GEANT4 model to assess the access growth in emittance that might have been caused by the unmatched beam and plasma conditions, apart from the scattering contribution. The preliminary prediction is an initial emittance of $1\,$mm$\,$mrad and a plasma density of $1\times10^{18}\,$m$^{-3}$ should provide the matching between beam and plasma and hence the minimum emittance. Studies towards a numerical demonstration of the phenomenon are ongoing. 
\section{Conclusions and Outlook}
As an advanced accelerating technique plasma wakefield acceleration has ever-increasing prospects. Therefore, the potential issues of the scheme must be assessed carefully. This study was initiated to seek out the impact of the interaction of a witness beam with the surrounding plasma formed to provide acceleration. The emittance growth induced in the beam via beam-gas elastic scattering was studied, numerically. It has been analytically calculated for the given parameters that the beam-plasma interaction can cause an emittance growth of $0.014\,nm$ after $500\,m$ travel within the plasma due to multiple Coulomb scattering. In order to analyse the total beam emittance, the sensitivity of our current GEANT4 model for the unmatched beam and plasma parameters are under study. 
\section{Acknowledgements}
This work was supported by the Cockcroft Institute Core Grant and STFC.

%
%
\raggedend


\begin{thebibliography}{9}   
\bibitem{Tajima} T. Tajima and J.M. Dawson, Phys. Rev. Lett. 43, 267 (1979).
\bibitem{Mangles} S.P.D. Mangles et al., Nature 431, 535 (2004).
\bibitem{Geddes} C.G.R. Geddes et al., Nature 431, 538 (2004).
\bibitem{Leemans1} W.P. Leemans, et al., Nat. Phys. 2, 696 (2006).
\bibitem{Leemans2} W. P. Leemans et al., Phys. Rev. Lett. 113, 245002 (2014).
\bibitem{Wang} X. Wang et al., Nature Commun. 4,1988 (2013).
\bibitem{Blumenfeld} I. Blumenfeld et al., Nature 445, 741 (2007).
\bibitem{base_theory}  B. W. Montague, AIP Conf. Proc. 07, 130(1):146-155 (1985).
\bibitem{Kirby} N. Kirby et al., Proc. of PAC07, p.3097 (2007).
\bibitem{geant4_1}  S. Agostinelli et al., NIM A 506, 250-303 (2003).
\bibitem{geant4_2}  J. Allison et al., IEEE Trans. on Nucl. Sci. 53, 1, p270 (2006).
\bibitem{LCODE} K.V.Lotov, PRSTAB, v.6, p.061301 (2003).
\bibitem{Oz}  E. Oz and P. Muggli, NIM A740, 197-202 (2014).
\bibitem{unmatched} C.E. Clayton et al., PRL, 15, 154801 (2002).
\bibitem{matched} P. Muggli, PRL, 93, 1014802 (2004) 

\end{thebibliography}
\end{document}